\documentclass[conference,a4paper]{APSIPA2018}
\usepackage{multirow}
\usepackage{amsmath}
\usepackage[psamsfonts]{amssymb}
\usepackage{amsxtra}
\usepackage{threeparttable}
\usepackage{booktabs}
\usepackage{graphicx}

\begin{document}

\title{Attention Based Fully Convolutional Network for Speech Emotion Recognition}

\author{%
\authorblockN{%
Yuanyuan Zhang, Jun Du, Zirui Wang, Jianshu Zhang, Yanhui Tu
}
\authorblockA{%
National Engineering Laboratory for Speech and Language Information Processing,}
\authorblockA{%
University of Science and Technology of China, Hefei, China \\
\{zyuan, cs211, xysszjs, tuyanhui\}@mail.ustc.edu.cn, jundu@ustc.edu.cn}
}

\maketitle
\thispagestyle{empty}

\begin{abstract}
Speech emotion recognition is a challenging task for three main reasons: 1) human emotion is abstract, which means it is hard to distinguish; 2) in general, human emotion can only be detected in some specific moments during a long utterance; 3) speech data with emotional labeling is usually limited. In this paper, we present a novel attention based fully convolutional network for speech emotion recognition. We employ fully convolutional network as it is able to handle variable-length speech, free of the demand of segmentation to keep critical information not lost. The proposed attention mechanism can make our model be aware of which time-frequency region of speech spectrogram is more emotion-relevant. Considering limited data, the transfer learning is also adapted to improve the accuracy. Especially, it's interesting to observe obvious improvement obtained with natural scene image based pre-trained model. Validated on the publicly available IEMOCAP corpus, the proposed model outperformed the state-of-the-art methods with a weighted accuracy of 70.4\% and an unweighted accuracy of 63.9\% respectively.
\end{abstract}

\section{Introduction}
Emotions play an important role in human communications~\cite{cowie2001emotion} and successfully detecting the emotion states is helpful to improve the efficiency of  human-computer interaction. For instance, in call centers, tracking customers' emotion states can be useful for quality measurement~\cite{burkhardt2006detecting} and the calls from angry customers can therefore be assigned to experienced agents. Speech is one of the communication channels that emotions could have serious influence on. Technically, emotions affect both the voice characteristics and linguistic content. In this study, we focus on the change of voice characteristics to recognize the underlying emotions in speech.

\begin{figure}[!htp]
\centering
\includegraphics[width=0.3\linewidth]{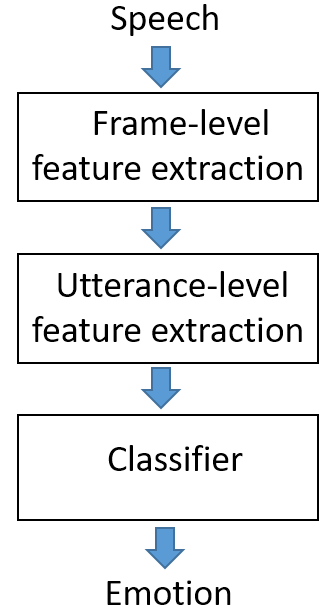}
\caption{The traditional speech emotion recognition system.}
\label{fig_tra_process}
\end{figure}

Speech emotion recognition (SER) has been an active research field for decades~\cite{vinola2015survey,el2011survey,chandrasekar2014automatic,koolagudi2012emotion}. We demonstrate the architecture of  traditional approaches for SER in Figure~\ref{fig_tra_process}. First, acoustic features which are believed to incorporate the information of human emotions are extracted from raw speech waveform frame by frame. The features include pitch, voicing probability, energy, etc. Then various statistical functions (e.g. mean, max, linear regression coefficients, etc.) are applied to the frame-level features. And the outputs are concatenated as a feature vector to represent the whole utterance. Finally, the utterance feature vector is fed to the classifier. There are many classification models that have been used~\cite{vinola2015survey,el2011survey,chandrasekar2014automatic,koolagudi2012emotion}, with support vector machine (SVM) being one of the most popular choices.

Recently, deep learning methods have been introduced to this field. In~\cite{stuhlsatz2011deep}, deep neural network (DNN) was used on the top of traditional utterance-level features and achieved a significant improvement on the accuracy compared with conventional classifiers. \cite{han2014speech} used DNN to learn the short-term acoustic features, followed by traditional statistical functions to construct utterance-level features and the extreme learning machine (ELM) was used as the classifier. In~\cite{satt2017efficient}, the state-of-the-art result was reported by using both convolutional and recurrent layers to directly learn the mapping from speech spectrogram to the corresponding emotion state. In~\cite{satt2017efficient}, the speech spectrogram must be segmented into pieces or zero-padded to a fixed size to satisfy the requirement of convolutional neural network (CNN). Each sub-utterance was assigned the emotion label of the corresponding whole utterance. And during the testing procedure, the prediction of the whole utterance was evaluated by averaging the posterior probabilities of all sub-utterances. However, it is not quite reasonable to assume that each sub-sentence within a whole sentence represents the overall emotion. In addition, the speech continuity could be destroyed by segmentation which made the system more difficult to catch the whole process of emotion changing from rise to fall.

To solve this problem, in this study, the fully convolutional neural network (FCN) is adopted to handle variable-length speech, free of the demand of segmentation to keep critical information not lost. In addition, attention mechanism has shown its efficiency especially in encoder-decoder models~\cite{bahdanau2014neural, luong2015effective, zhang2017watch}, which is employed to guide the decoder to know which parts of the outputs of the encoder are more important. Specific to classification models, a self-attention mechanism has been proposed and it is designed to tell the classifier which parts of the input are more relevant to the output classes. In~\cite{yang2016hierarchical, lin2017structured}, the self-attention was used to extract sentence embedding for semantic analysis. In~\cite{mirsamadi2017automatic}, the authors used the self-attention mechanism on SER, enabling the network to focus on emotional salient of an utterance. The encoder they adopted is long short-term memory (LSTM). Considering many irrelevant signals are mixed with speech signals, we adopt attention mechanism with FCN to achieve 2D attention visualization on top of spectrograms rather than 1D attention visualization only on the time axis in~\cite{mirsamadi2017automatic}.

Another problem is the that speech data with emotional labeling is usually hard to collect. Transfer learning is a useful method to solve the current task with the help of the knowledge obtained from related problems~\cite{bahdanau2014neural}, which can be operated as finetuning the parameters of network from a pre-trained model, It has been widely used when the training data is insufficient~\cite{girshick2014rich, donahue2014decaf, long2015fully, koller2016deep}, especially when the model is based on CNN. In this paper, we present an interesting observation, i.e., an obvious improvement on SER can be obtained with natural scene image based pre-trained model. It worth noting that speech signal is very different from image.

The remainder of the paper is organized as follows. In Section 2, we first introduce the proposed architecture. In Section 3, we report and analyze experiment results. Finally we summarize our work and present conclusions in Section 4.

\section{The Proposed Architecture}
\begin{figure}[!htp]
\centering
\includegraphics[width=0.6\linewidth]{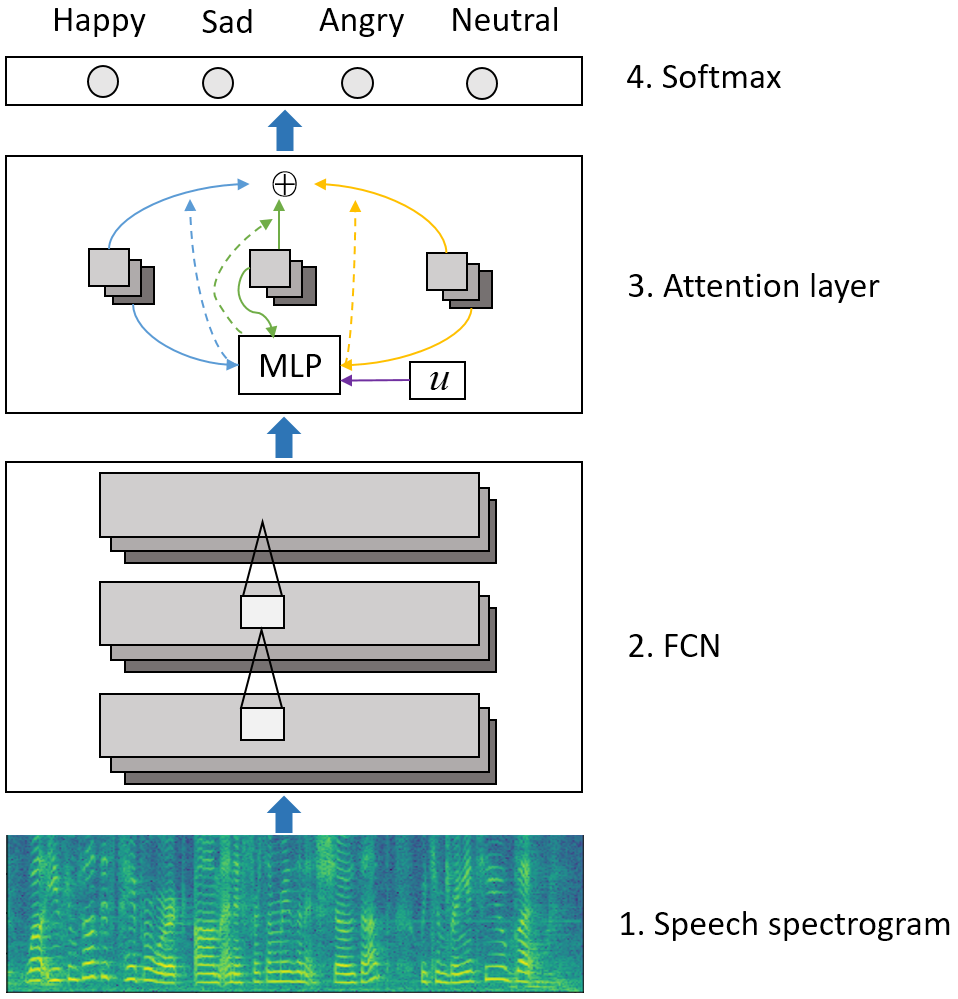}
\caption{The overall architecture of an attention based fully convolutional neural network.}
\label{fig_overall_arch}
\end{figure}

In this paper, we propose a novel attention based fully convolutional neural network. The input of the model is also the spectrogram. But inherently unlike \cite{satt2017efficient}, we do not need to segment spectrograms into pieces or pad them to a fixed shape. The FCN is able to handle spectrogram with variable sizes. The overall architecture is shown in Figure~\ref{fig_overall_arch}. The FCN encodes the spectrogram into a high-level representation while the attention mechanism impels the remaining sub-layers of the model to focus on specific time-frequency regions of the input spectrogram. All components of the system can be optimized jointly.

\begin{figure}[!ht]
\centering
\includegraphics[width=0.5\linewidth]{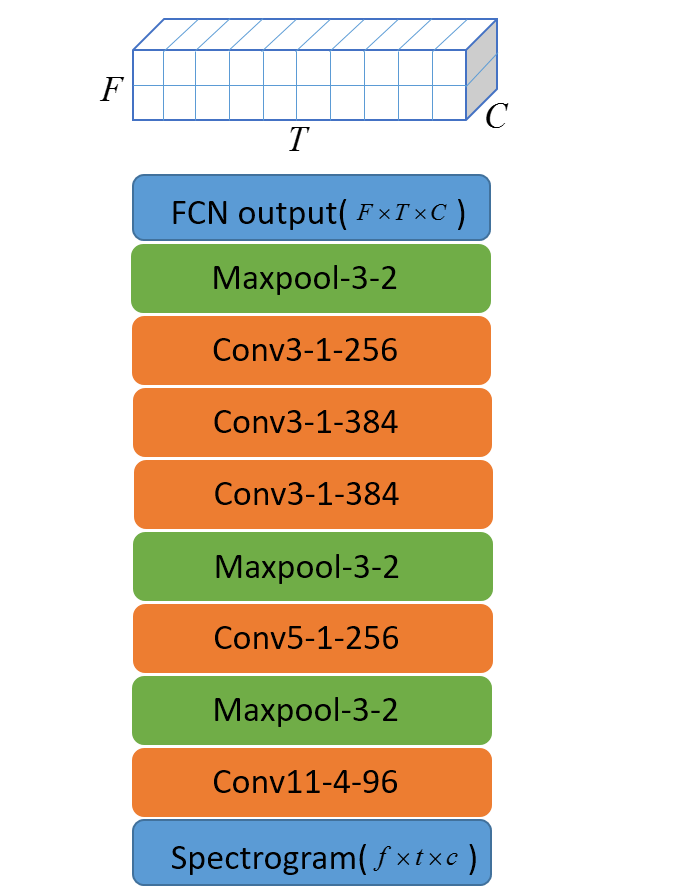}
\caption{The AlexNet based FCN configurations. The convolutional layer parameters are denoted as  ``Conv(kernel size)-[stride size]-[number of channels]''. The maxpooling layer parameters are denoted as ``Maxpool-[kernel size]-[stride size]''. For brevity, the local response normalization layer and ReLU activation function is not shown.}
\label{encoder}
\end{figure}

\subsection{Fully convolutional network}

CNN has been widely used for deep learning, which does not require traditional handcrafted feature extraction and it has been proved that CNN based system can obtain a comparable or even better accuracy compared with the traditional systems on the SER task  \cite{mirsamadi2017automatic, trigeorgis2016adieu, aldeneh2017using}. The basic components of CNN  are convolution, pooling and activation layers. The convolutional layer is determined by the number of input channels, the number of output feature maps, the kernel size and stride. Each kernel can be considered as a filter whose size is usually much smaller than the input. Hence, a kernel operates on a local region of input rather than the whole feature map. The locations that connect to higher layers are called receptive fields. On a given feature map, the kernel weights are shared to detect certain feature in different locations and to reduce the complexity of network. The pooling layers usually conduct an average or max pooling operation to remove noise and extract robust features. The activation layers are actually element-wise nonlinear functions \cite{zhang2017watch}.

The typical CNNs, including AlexNet \cite{krizhevsky2012imagenet}, Oxford VGGNet \cite{simonyan2014very}, and ResNet \cite{he2016deep} take fixed-size input. Inspired by \cite{long2015fully},  we turn the AlexNet into a fully convolutional network by simply removing its fully connected layers. And then it is used as our encoder, which is shown in Figure~\ref{encoder}. All the convolution layers are followed by a ReLU activation function, and the first two convolution layers are equipped with a local response normalization layer. We also directly adapt the VGGnet to classify emotion states, but it yields a lower accuracy than AlexNet due to the limited training data.

Assuming that the output of FCN encoder is a 3-dimensional array of size $F\times T\times C$,where the F and T correspond to the frequency and time domains of spectrogram and C is channel size. We can consider the output as a variable-length grid of $L$ elements, $L=F\times T$. Each of the elements is a C-dimensional vector corresponding to a region of speech spectrogram, represented as $\boldsymbol{a}_{i}$.
\begin{equation}
\label{FCN_output}
\boldsymbol{A} =\{ \boldsymbol{a}_{1},\cdots,\boldsymbol{a}_{L}\}  ,  \boldsymbol{a}_{i} \in \mathbb{R}^C
\end{equation}

\subsection{Attention layer}
Intuitively, not all time-frequency units contribute equally to the emotion state of the whole utterance, i.e., not all the element vectors of set $\boldsymbol{A}$ contribute equally to the emotion state. Hence, we introduce attention mechanism to extract the elements that are important to the emotion of the utterance and aggregate those element arrays to form an utterance emotion vector. We use the following formulas to realize this idea:
\begin{equation}
\label{attention_trans_inner}
e_{i} = \boldsymbol{u}^{T} \tanh(\boldsymbol{W}\boldsymbol{a}_{i}+\boldsymbol{b})
\end{equation}
\begin{equation}
\label{scale_softmax}
\alpha_{i}=\frac{\exp(\lambda e_i)}{\sum_{k=1}^{L}\exp(\lambda e_k)}
\end{equation}
\begin{equation}
\label{weighted_sum}
\boldsymbol{c} = \sum_{i=1}^{L}\alpha_{i}\boldsymbol{a}_{i}
\end{equation}

That is, first we feed the annotation $\boldsymbol{a}_{i}$ through a multi-layer-perceptron (MLP) layer with the $\tanh$ as the non-linear activation function to obtain a new representation of $\boldsymbol{a}_{i}$. Then we measure the importance weight, $e_{i}$, of the $\boldsymbol{a}_{i}$ by the inner product between this new vector and the learnable vector $\boldsymbol{u}$. After that, the normalized importance weight $\alpha_{i}$ is calculated through the softmax function. Finally, the utterance emotion vector $\boldsymbol{c}$ is computed as the weighted sum of set $\boldsymbol{A}$ with importance weights. $\lambda$ is a scale factor which controls the uniformity of the importance weights of the annotation vectors. $\lambda$ ranges between 0 and 1. If $\lambda=1$, the scaled-softmax becomes the commonly used softmax function. If $\lambda=0$, the importance weights will be a uniform distribution on the set $\boldsymbol{A}$, which means all the time-frequency units have the same importance weights for the final utterance emotion vector. In this study, $\lambda=0.3$ is used for the balance.

\section{Experiments}
\subsection{Database and feature extraction}
We validate our systems on the IEMOCAP database \cite{busso2008iemocap}, one of the widely used databases on speech emotion recognition. The IEMOCAP corpus comprises five sessions, each of which includes labeled emotional speech utterances from recordings of dialogs between two actors. There is no actor overlapping between these sessions. To be comparable with \cite{satt2017efficient}, we utilize the database in the same way:
\begin{itemize}
\item The IEMOCAP database contains scripted and improvised dialogs. We only use improvised data.
\item We use the speech utterances from four emotion categories, i.e., happy, sad, angry and neutral.
\item We implement a five-fold cross validation. In each fold, the data from four sessions is used for model training, and the data from the remaining session is splited: one actor for validation and the other one as the testing set.
\end{itemize}

The experiments only apply the raw spectrogram as the input, the spectrogram extraction process is consistent with \cite{satt2017efficient}: First, a sequence of overlapping Hamming windows are applied to the speech waveform, with window shift set to 10 msec, and window size set to 40 msec. Then, for each frame we calculate a discrete Fourier transform (DFT) of length 800. Finally the 200-dimensional low-frequency part of the spectrogram is used as the input. Please note that in \cite{satt2017efficient}, the Hamming window size of 20 msec is used, and the authors concluded that the size of 20 msec is better. In our study, we set the window size to 40 msec and achieve a higher accuracy.

\subsection{Evaluation metric}
The IEMOCAP database is imbalanced with respect to the emotional classes. So we adopt both the weighted accuracy (WA) and the unweighted accuracy (UA) as the metric:
\begin{itemize}
\item Weighted accuracy - the overall accuracy across all utterances of the testing set.
\item Unweighted accuracy - the average of accuracies across all the classes.
\end{itemize}

\subsection{Experiment results and analysis}

\begin{table}[t]
  \caption{The accuracy comparison of AlexNet and VGGNet-16 with random initialization or finetuning.}
  \label{tab:finetune_or_random}
  \centering
  \newcommand{\tabincell}[2]{\begin{tabular}{@{}#1@{}}#2\end{tabular}}
  \begin{tabular}{lcc}
   \toprule
   \textbf{System}& \tabincell{c}{\textbf{Weighted} \\ \textbf{Accuracy}}& \tabincell{c}{\textbf{Unweighted} \\ \textbf{Accuracy}}\\
   \midrule
    AlexNet Random-init & $66.5\%$ & $54.8\%$ \\
   AlexNet Finetuning & $\boldsymbol{67.9\%}$ & $\boldsymbol{57.3\%}$ \\
    VGGNet-16 Random-init & $65.3\%$ & $54.8\%$ \\
   VGGNet-16 Finetuning & $66.8\%$ & $ 56.7\%$ \\
   \bottomrule
  \end{tabular}
\end{table}

First, we directly adapt AlexNet and VGGNet-16 to classify. The only difference is that there are 4 nodes in softmax layer. The utterances are split or padded to fixed-length sub-utterances by using the same method in \cite{satt2017efficient}. During the testing procedure, the posterior probabilities are the average of the all sub-utterances respectively. Considering the limited data, we compare the networks with random initialization and the pre-trained networks based on ImageNet dataset \cite{russakovsky2015imagenet}.

Table~\ref{tab:finetune_or_random} summarizes the results of AlexNet and VGGNet-16 with different initializations. It's interesting to observe the pre-trained neural networks (NNs) always outperform the NNs with random initialization. It's worth noting that the speech signal is very different from image. The only explanation is the pre-trained NNs have been empowered to detect some certain structures so that they can be more easily trained. By comparing the first row and third row (or comparing the second row and fourth row), we demonstrate that the AlexNet outperforms the VGGNet-16 on this task. We think the lack of sufficient training speech data is one main reason. Based on these results, our FCN model directly uses AlexNet, excluding its full connected layers. And we initialize FCN by using the pre-trained parameters.

\begin{table}[t]
  \caption{The accuracy comparison between FCN based attention model and the other systems.}
  \label{tab:our_vs_other}
  \centering
  \newcommand{\tabincell}[2]{\begin{tabular}{@{}#1@{}}#2\end{tabular}}
  \begin{tabular}{lcc}
   \toprule
   \textbf{System}& \tabincell{c}{\textbf{Weighted} \\ \textbf{Accuracy}}& \tabincell{c}{\textbf{Unweighted} \\ \textbf{Accuracy}}\\
   \midrule
   Our FCN+Attention & $\boldsymbol{70.4\%}$ & $\boldsymbol{63.9\%}$  \\
   CNN+LSTM Model1 in \cite{satt2017efficient} & $68.8\%$ & $59.4\%$ \\
   CNN+LSTM Model2 in \cite{satt2017efficient} & $67.3\%$ & $62.0\%$ \\
   \bottomrule
  \end{tabular}
\end{table}

\begin{figure*}[t]
\centering
\includegraphics[width=0.9\linewidth]{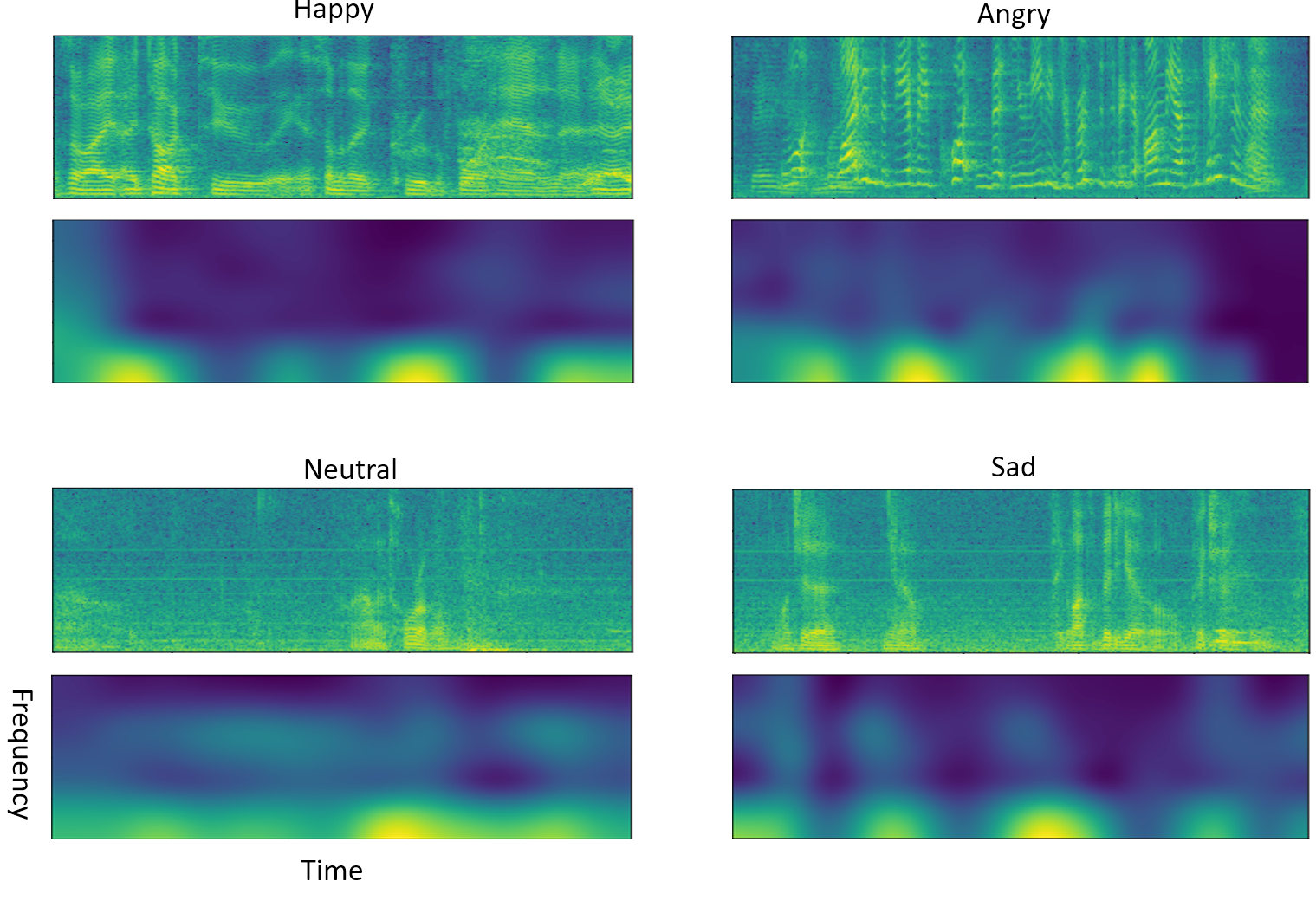}
\caption{The 2D-attention weights of FCN model for 4 examples in different emotion categories. Top: The spectrogram. Bottom: The 2D-attention weights figure of spectrogram. Each point of the figure corresponds to the point of spectrogram in the same location and the brighter color represents the larger weight.}
\label{fig_d2_vs_d1}
\end{figure*}

The published state-of-the-art results using the IEMOCAP corpus are given in \cite{satt2017efficient}. We list their two best models, i.e., CNN+LSTM Model1 and CNN+LSTM Model2 in Table~\ref{tab:our_vs_other}. Model1 is a CNN-LSTM model while Model2 is trained based on Model1 in order to improve the unweighted accuracy. The attention based FCN model is trained just by one step. And compared with the best results in both Model1 and Model2, our attention based FCN model achieves $1.6\%$ and $1.9\%$ absolute improvements on WA and UA, respectively.

To explain why the improvement can be gained from our proposed approach, we plot the 2D-attention weights of FCN model for 4 test examples of different emotion categories in Figure~\ref{fig_d2_vs_d1}. The Figure~\ref{fig_d2_vs_d1} illustrates that the attention weights of the non-speech frames are quite small which indicates that the voice activation detection is implicitly implemented and the information from non-speech frames are ignored by the attention mechanism automatically. Besides, the time-frequency units of spectrogram are assigned different weights based on the degrees they are relevant to emotion states. That explains why the attention weights are also small on parts of the voice frames. And the attention weights always have small values in high frequency areas, which is consistent with the common sense that the information of speech is mainly contained in the low frequency area. Actually, the bright area extends from low frequency to high frequency with a decreasing brightness. This indicates our 2D-attention mechanism has detected the emotional segment successfully and assigned decreasing weights from low to high frequency bands. The 2D-attention mechanism is able to scan the spectrogram not only in the time domain but also in the frequency domain.

\section{Conclusions}
We demonstrated that the CNN architectures designed for visual recognition can be directly adapted for speech emotion recognition. Besides, it's interesting to see the transfer learning can build a solid bridge between natural image and speech signal. Finally, we proposed an attention based FCN model. Our model is able to handle utterances with variable lengths and the attention mechanism empowers the network to focus on emotionally salient regions of spectrogram. Our system achieves beyond the state-of-the-art accuracy on the benchmark dataset IEMOCAP.

\bibliographystyle{IEEEtran}

\bibliography{mybib}

\begin{thebibliography}{10}
\providecommand{\url}[1]{#1}
\csname url@samestyle\endcsname
\providecommand{\newblock}{\relax}
\providecommand{\bibinfo}[2]{#2}
\providecommand{\BIBentrySTDinterwordspacing}{\spaceskip=0pt\relax}
\providecommand{\BIBentryALTinterwordstretchfactor}{4}
\providecommand{\BIBentryALTinterwordspacing}{\spaceskip=\fontdimen2\font plus
\BIBentryALTinterwordstretchfactor\fontdimen3\font minus
  \fontdimen4\font\relax}
\providecommand{\BIBforeignlanguage}[2]{{%
\expandafter\ifx\csname l@#1\endcsname\relax
\typeout{** WARNING: IEEEtran.bst: No hyphenation pattern has been}%
\typeout{** loaded for the language `#1'. Using the pattern for}%
\typeout{** the default language instead.}%
\else
\language=\csname l@#1\endcsname
\fi
#2}}
\providecommand{\BIBdecl}{\relax}
\BIBdecl

\bibitem{cowie2001emotion}
R.~Cowie, E.~Douglas-Cowie, N.~Tsapatsoulis, G.~Votsis, S.~Kollias, W.~Fellenz,
  and J.~G. Taylor, ``Emotion recognition in human-computer interaction,''
  \emph{IEEE Signal processing magazine}, vol.~18, no.~1, pp. 32--80, 2001.

\bibitem{burkhardt2006detecting}
F.~Burkhardt, J.~Ajmera, R.~Englert, J.~Stegmann, and W.~Burleson, ``Detecting
  anger in automated voice portal dialogs,'' in \emph{Ninth International
  Conference on Spoken Language Processing}, 2006.

\bibitem{vinola2015survey}
C.~Vinola and K.~Vimaladevi, ``A survey on human emotion recognition
  approaches, databases and applications,'' \emph{ELCVIA Electronic Letters on
  Computer Vision and Image Analysis}, vol.~14, no.~2, pp. 24--44, 2015.

\bibitem{el2011survey}
M.~El~Ayadi, M.~S. Kamel, and F.~Karray, ``Survey on speech emotion
  recognition: Features, classification schemes, and databases,'' \emph{Pattern
  Recognition}, vol.~44, no.~3, pp. 572--587, 2011.

\bibitem{chandrasekar2014automatic}
P.~Chandrasekar, S.~Chapaneri, and D.~Jayaswal, ``Automatic speech emotion
  recognition: A survey,'' in \emph{Circuits, Systems, Communication and
  Information Technology Applications (CSCITA), 2014 International Conference
  on}.\hskip 1em plus 0.5em minus 0.4em\relax IEEE, 2014, pp. 341--346.

\bibitem{koolagudi2012emotion}
S.~G. Koolagudi and K.~S. Rao, ``Emotion recognition from speech: a review,''
  \emph{International journal of speech technology}, vol.~15, no.~2, pp.
  99--117, 2012.

\bibitem{stuhlsatz2011deep}
A.~Stuhlsatz, C.~Meyer, F.~Eyben, T.~Zielke, G.~Meier, and B.~Schuller, ``Deep
  neural networks for acoustic emotion recognition: raising the benchmarks,''
  in \emph{Acoustics, speech and signal processing (ICASSP), 2011 IEEE
  international conference on}.\hskip 1em plus 0.5em minus 0.4em\relax IEEE,
  2011, pp. 5688--5691.

\bibitem{han2014speech}
K.~Han, D.~Yu, and I.~Tashev, ``Speech emotion recognition using deep neural
  network and extreme learning machine,'' in \emph{Fifteenth Annual Conference
  of the International Speech Communication Association}, 2014.

\bibitem{satt2017efficient}
A.~Satt, S.~Rozenberg, and R.~Hoory, ``Efficient emotion recognition from
  speech using deep learning on spectrograms,'' \emph{Proc. Interspeech 2017},
  pp. 1089--1093, 2017.

\bibitem{bahdanau2014neural}
D.~Bahdanau, K.~Cho, and Y.~Bengio, ``Neural machine translation by jointly
  learning to align and translate,'' \emph{arXiv preprint arXiv:1409.0473},
  2014.

\bibitem{luong2015effective}
M.-T. Luong, H.~Pham, and C.~D. Manning, ``Effective approaches to
  attention-based neural machine translation,'' \emph{arXiv preprint
  arXiv:1508.04025}, 2015.

\bibitem{zhang2017watch}
J.~Zhang, J.~Du, S.~Zhang, D.~Liu, Y.~Hu, J.~Hu, S.~Wei, and L.~Dai, ``Watch,
  attend and parse: An end-to-end neural network based approach to handwritten
  mathematical expression recognition,'' \emph{Pattern Recognition}, vol.~71,
  pp. 196--206, 2017.

\bibitem{yang2016hierarchical}
Z.~Yang, D.~Yang, C.~Dyer, X.~He, A.~Smola, and E.~Hovy, ``Hierarchical
  attention networks for document classification,'' in \emph{Proceedings of the
  2016 Conference of the North American Chapter of the Association for
  Computational Linguistics: Human Language Technologies}, 2016, pp.
  1480--1489.

\bibitem{lin2017structured}
Z.~Lin, M.~Feng, C.~N.~d. Santos, M.~Yu, B.~Xiang, B.~Zhou, and Y.~Bengio, ``A
  structured self-attentive sentence embedding,'' \emph{arXiv preprint
  arXiv:1703.03130}, 2017.

\bibitem{mirsamadi2017automatic}
S.~Mirsamadi, E.~Barsoum, and C.~Zhang, ``Automatic speech emotion recognition
  using recurrent neural networks with local attention,'' in \emph{Acoustics,
  Speech and Signal Processing (ICASSP), 2017 IEEE International Conference
  on}.\hskip 1em plus 0.5em minus 0.4em\relax IEEE, 2017, pp. 2227--2231.

\bibitem{girshick2014rich}
R.~Girshick, J.~Donahue, T.~Darrell, and J.~Malik, ``Rich feature hierarchies
  for accurate object detection and semantic segmentation,'' in
  \emph{Proceedings of the IEEE conference on computer vision and pattern
  recognition}, 2014, pp. 580--587.

\bibitem{donahue2014decaf}
J.~Donahue, Y.~Jia, O.~Vinyals, J.~Hoffman, N.~Zhang, E.~Tzeng, and T.~Darrell,
  ``Decaf: A deep convolutional activation feature for generic visual
  recognition,'' in \emph{International conference on machine learning}, 2014,
  pp. 647--655.

\bibitem{long2015fully}
J.~Long, E.~Shelhamer, and T.~Darrell, ``Fully convolutional networks for
  semantic segmentation,'' in \emph{Proceedings of the IEEE conference on
  computer vision and pattern recognition}, 2015, pp. 3431--3440.

\bibitem{koller2016deep}
O.~Koller, O.~Zargaran, H.~Ney, and R.~Bowden, ``Deep sign: hybrid cnn-hmm for
  continuous sign language recognition,'' in \emph{Proceedings of the British
  Machine Vision Conference 2016}, 2016.

\bibitem{trigeorgis2016adieu}
G.~Trigeorgis, F.~Ringeval, R.~Brueckner, E.~Marchi, M.~A. Nicolaou,
  B.~Schuller, and S.~Zafeiriou, ``Adieu features? end-to-end speech emotion
  recognition using a deep convolutional recurrent network,'' in
  \emph{Acoustics, Speech and Signal Processing (ICASSP), 2016 IEEE
  International Conference on}.\hskip 1em plus 0.5em minus 0.4em\relax IEEE,
  2016, pp. 5200--5204.

\bibitem{aldeneh2017using}
Z.~Aldeneh and E.~M. Provost, ``Using regional saliency for speech emotion
  recognition,'' in \emph{Acoustics, Speech and Signal Processing (ICASSP),
  2017 IEEE International Conference on}.\hskip 1em plus 0.5em minus
  0.4em\relax IEEE, 2017, pp. 2741--2745.

\bibitem{krizhevsky2012imagenet}
A.~Krizhevsky, I.~Sutskever, and G.~E. Hinton, ``Imagenet classification with
  deep convolutional neural networks,'' in \emph{Advances in neural information
  processing systems}, 2012, pp. 1097--1105.

\bibitem{simonyan2014very}
K.~Simonyan and A.~Zisserman, ``Very deep convolutional networks for
  large-scale image recognition,'' \emph{arXiv preprint arXiv:1409.1556}, 2014.

\bibitem{he2016deep}
K.~He, X.~Zhang, S.~Ren, and J.~Sun, ``Deep residual learning for image
  recognition,'' in \emph{Proceedings of the IEEE conference on computer vision
  and pattern recognition}, 2016, pp. 770--778.

\bibitem{busso2008iemocap}
C.~Busso, M.~Bulut, C.-C. Lee, A.~Kazemzadeh, E.~Mower, S.~Kim, J.~N. Chang,
  S.~Lee, and S.~S. Narayanan, ``Iemocap: Interactive emotional dyadic motion
  capture database,'' \emph{Language resources and evaluation}, vol.~42, no.~4,
  p. 335, 2008.

\bibitem{russakovsky2015imagenet}
O.~Russakovsky, J.~Deng, H.~Su, J.~Krause, S.~Satheesh, S.~Ma, Z.~Huang,
  A.~Karpathy, A.~Khosla, M.~Bernstein \emph{et~al.}, ``Imagenet large scale
  visual recognition challenge,'' \emph{International Journal of Computer
  Vision}, vol. 115, no.~3, pp. 211--252, 2015.

\end{thebibliography}
\end{document}